\documentclass[final,1p,times]{elsarticle}
\usepackage{graphicx}
\usepackage{amssymb}
\usepackage[section]{placeins}

\usepackage[font=footnotesize]{caption}
\usepackage[normalem]{ulem}
\usepackage{color}
\usepackage{amsmath}
\usepackage{bm}
\usepackage{amsfonts}
\makeatletter
\def\amsbb{\use@mathgroup \M@U \symAMSb}
\makeatother
\usepackage{bbold}

\biboptions{numbers,sort&compress}

\renewcommand\({\left(}
\renewcommand\){\right)}
\renewcommand\[{\left[}
\renewcommand\]{\right]}

\newcommand{\bp}{{\bf p}}
\newcommand{\bv}{{\bf v}}
\newcommand{\br}{{\bf r}}
\newcommand{\bk}{{\bf k}}
\newcommand{\GF}{G_{\rm F}}
\newcommand{\I}{{\rm i}}
\newcommand{\half}{{\textstyle\frac{1}{2}}}

\begin{document}

\begin{frontmatter}


 %\tnotetext[label1]{}
 %\author{Name\corref{cor1}\fnref{label2}}
 %\ead{email address}
 %\fntext[label2]{}
 %\cortext[cor1]{}
 %\address{Address\fnref{label2}}
 %\fntext[label3]{}

\title{Collective neutrino flavor conversion: Recent developments}

\author[label1]{Sovan Chakraborty}
\author[label2]{Rasmus Hansen}
\author[label1]{Ignacio Izaguirre}
\author[label1]{Georg Raffelt}
\address[label1]{\it Max-Planck-Institut f{\"u}r Physik (Werner-Heisenberg-Institut),
  F{\"o}hringer Ring 6, 80805 M{\"unchen}, Germany}
\address[label2]{\it Max-Planck-Institut f{\"u}r Kernphysik,
  Saupfercheckweg 1, 69117 Heidelberg, Germany}

\begin{abstract}
Neutrino flavor evolution in core-collapse supernovae, neutron-star
mergers, or the early universe is dominated by neutrino-neutrino
refraction, often spawning ``self-induced flavor conversion,'' i.e.,
shuffling of flavor among momentum modes. This effect is driven by
collective run-away modes of the coupled ``flavor oscillators'' and can
spontaneously break the initial symmetries such as axial symmetry,
homogeneity, isotropy, and even stationarity. Moreover, the growth rates of
unstable modes can be of the order of the neutrino-neutrino interaction energy
instead of the much smaller vacuum oscillation frequency: self-induced
flavor conversion does not always require neutrino masses. We illustrate
these newly found phenomena in terms of simple toy models. What happens in
realistic astrophysical settings is up to speculation at present.
\end{abstract}

\begin{keyword}
Neutrino oscillations \sep supernova\end{keyword}
\end{frontmatter}

\section{Introduction}
\label{sec:introduction}

Neutrino dispersion in matter strongly modifies the flavor evolution caused
by their masses and mixing parameters \cite{Wolfenstein:1977ue,
Mikheev:1986gs, Kuo:1989qe}. Moreover, in dense astrophysical environments,
notably in core-collapse supernovae, merging neutron stars, or the early
universe, neutrinos themselves provide a large refractive effect. Flavor
evolution implies that the ``background medium'' also evolves, i.e., neutrino
flavor evolution feeds back unto itself \cite{Pantaleone:1992eq, Sigl:1992fn}
and, among other effects, can produce collective run-away modes in flavor
space. One consequence is ``self-induced flavor conversion,'' meaning that
some modes of the neutrino mean field swap flavor with other modes
\cite{Samuel:1993uw, Kostelecky:1993dm, Samuel:1995ri, Sawyer:2005jk,
Duan:2005cp, Hannestad:2006nj, Duan:2007mv}. To lowest order,
neutrino-neutrino interactions are flavor blind \cite{Mirizzi:2009td}, so
collective effects alone do not change the global flavor content of the
ensemble. Yet the reshuffling among modes can effectively engender flavor
equilibration. The simplest example is a gas of equal densities of $\nu_e$
and $\bar\nu_e$ that would turn to an equal $\nu$-$\bar\nu$ mixture of all
flavors, yet the overall flavor lepton numbers remain zero from beginning to
end \cite{Raffelt:2007yz}.

The neutrino mean field is described by flavor matrices
$\varrho(t,{\br},{\bp})$ with elements of the type \smash{$\langle
a_i^\dagger a_j\rangle$} in terms of creation and annihilation operators with
flavor index $i$. The diagonal elements are occupation numbers, the
off-diagonal ones represent flavor correlations. The seven-dimensional phase
space $(t,{\bf r},{\bf p})$ is not tractable and was always reduced by
symmetry assumptions. For supernova neutrinos, one has usually assumed
stationary solutions which depend only on radial distance and, in momentum
space, on energy and zenith angle, reducing the problem to three dimensions.
The emission region was often modeled as a ``neutrino bulb,'' meaning an
emitting surface (``neutrino sphere''), where neutrinos of all flavors emerge
with a blackbody-inspired zenith-angle distribution. Such models lead to
sharp spectral features (``spectral splits'') caused by flavor swaps between
different parts of the spectrum \cite{Duan:2006an, Raffelt:2007cb,
Fogli:2007bk, Duan:2010bg}. Being triggered by an instability which is
sensitive to the neutrino mass ordering, these effects seemed to offer an
opportunity to learn about the latter even for a very small $\Theta_{13}$
mixing angle \cite{Dasgupta:2008my}.

Meanwhile the situation has changed on several fronts. The mixing angle
$\Theta_{13}$ has been measured and is not very small, so the mass ordering
will be experimentally accessible. Moreover, our ideas about the flavor
evolution of supernova neutrinos had to be revised because it has dawned on
us that the earlier symmetry assumptions had constrained the solutions in
unphysical ways. Even within the bulb model of neutrino emission, axial
symmetry \cite{Raffelt:2013rqa, Raffelt:2013isa, Hansen:2014paa,
Mirizzi:2013wda, Chakraborty:2014nma}, homogeneity \cite{Mangano:2014zda,
Duan:2014gfa, Mirizzi:2015fva, Abbar:2015mca, Chakraborty:2015tfa}, and
stationarity \cite{Abbar:2015fwa, Dasgupta:2015iia} can be spontaneously
broken, completely changing the stability conditions. It has been speculated
that self-induced flavor conversion may commence in the decoupling region,
much deeper than the usual ``onset radius,'' and that flavor equilibration
could be a generic outcome instead of ordered spectral swaps.

Moreover, in a more realistic emission model, $\nu_e$, $\bar\nu_e$ and the
other species have different zenith-angle distributions. Surprisingly,
self-induced flavor conversion can then be ``fast'' in the sense that the
evolution speed is of the order of the neutrino-neutrino interaction energy
$\mu=\sqrt{2}\GF n_\nu$ instead of the much smaller vacuum oscillation
frequency $\omega=\Delta m^2/2E$ \cite{Sawyer:2005jk, Sawyer:2015dsa,
Chakraborty:2016lct}. While this phenomenon had been noted a long time ago
\cite{Sawyer:2005jk}, its significance had eluded much of the community. Fast
flavor conversion does not depend on neutrino masses,\footnote{The Physics
Nobel Prize 2015 was awarded ``for the discovery of neutrino oscillations,
which shows that neutrinos have mass.'' Ironically, self-induced flavor
conversion does not always depend on neutrino masses, although this
connection exists, of course, in the context of vacuum oscillations and
standard MSW conversion.} except perhaps for providing initial disturbances
to seed the run-away modes.\footnote{One may speculate that initial
disturbances could be provided even by quantum fluctuations of our classical
mean-field quantities. However, since neutrinos are known to mix and to have
small masses, in practice ordinary neutrino flavor oscillations are
guaranteed to provide disturbances even on the mean-field level.} This
counter-intuitive behavior owes to the character of self-induced flavor
conversion as an instability and to its nature of flavor shuffling among
modes which globally conserves flavor number.

Many studies of supernova neutrino flavor evolution were restricted to
two-flavor effects driven by the atmospheric mass difference and the small
mixing angle $\Theta_{13}$ because all effects driven by the solar mass
difference require larger distances. Moreover, the $\nu_\mu$, $\bar\nu_\mu$,
$\nu_\tau$ and $\bar\nu_\tau$ species, collectively called $\nu_x$, are
produced with similar spectra, so flavor conversion among them may not be
crucial. Even then, however, three-flavor effects can be important
\cite{Dasgupta:2007ws, EstebanPretel:2007yq, Duan:2008za, Dasgupta:2008cd,
Gava:2008rp, Dasgupta:2010ae, Dasgupta:2010cd} and also allow new
instabilities opened up by the additional degree of freedom
\cite{Friedland:2010sc}. Moreover, fast flavor conversion is independent of
$\Delta m^2$, so realistically one needs to include three flavors.

Most of these effects have been recognized only very recently and we have no
complete or even approximate picture yet of how flavor really evolves in
realistic astrophysical environments. In particular, realistic numerical
studies seem out of the question for the time being. If quick flavor
decoherence is a generic outcome, or at least if one can develop criteria
under which circumstances this will be the case, full-scale simulations may
not be needed. For the moment, it may be more fruitful to take a reductionist
approach and develop a better understanding of the various forms of behavior
shown by an interacting neutrino gas. In this spirit, we discuss the recent
developments here in the framework of simple toy models for illustration,
restricting ourselves to the method of linearized stability analysis
\cite{Banerjee:2011fj}.

We begin in Sec.~\ref{sec:eom} with the equations of motion for the
neutrino mean field on the refractive level and we explain the
geometric structure of our problem relevant for compact astrophysical
sources. In Sec.~\ref{sec:linearized} we linearize the equation in
terms of small off-diagonal elements of the $\varrho$ matrices,
assuming neutrinos were produced in weak-interaction states. In
Sec.~\ref{sec:beam} we set up our main ``gedanken experiment,'' a
neutrino gas consisting of a total of four modes: left- and
right-moving neutrinos and antineutrinos, i.e., colliding beams of
$\nu_e$ and $\bar\nu_e$. We study the simple cases of spontaneous
left-right symmetry breaking as well as the spontaneous breaking of
homogeneity in the form of unstable spatial Fourier modes.  In
Sec.~\ref{sec:fastflavor} we turn to examples with fewer initial
symmetries and show how fast flavor conversion appears, i.e.,
instabilities with growth rates of order $\mu$, the neutrino-neutrino
interaction energy, instead of the much smaller vacuum oscillation
frequency $\omega$.  In Sec.~\ref{sec:temporal} we comment on the
latest development: unstable time-varying modes from a stationary
source. In Sec.~\ref{sec:conclusions} we close with a brief summary
and outlook.

\section{Equations of motion and adopted geometry}
\label{sec:eom}

The flavor evolution of dense neutrino gases thus far has only been studied
on the refractive level, i.e., neutrinos were always taken to be freely
streaming without collisions. As a physical description one uses the neutrino
mean field, i.e., ``occupation numbers'' of the type $\langle a^\dagger_i
a_j\rangle$ in terms of the usual creation and annihilation operators. They
are assembled in the form of matrices in flavor space $\varrho(t,\br,E,\bv)$,
essentially providing classical phase-space densities. This mean-field
description ignores higher-order correlators, i.e., we work on the lowest
level of the BBGKY hierarchy and in addition we ignore spin and pairing
correlations \cite{Volpe:2013jgr, Vaananen:2013qja, Vlasenko:2013fja,
Serreau:2014cfa, Kartavtsev:2015eva}. It is most economical to include
antineutrinos in the equations as modes with negative energy so that $E$ can
be both positive and negative. The direction of motion is represented by the
velocity $\bv=\bp/|E|$ and we approximate $|\bv|=1$ for our
ultra-relativistic neutrinos. The diagonal elements of $\varrho$ for
antineutrinos (negative $E$ modes) are negative occupation numbers, which is
the ``flavor isospin convention.'' It is sometimes more intuitive to
represent antineutrinos by matrices $\bar\varrho$ of positive occupation
numbers and positive energies at the price of more cumbersome equations.

The space-time evolution of $\varrho(t,\br,E,\bv)$ is governed by free
streaming and by local flavor evolution in the form of a Liouville equation
\cite{Sigl:1992fn}
\begin{equation}\label{eq:EOM1}
\I(\partial_t+\bv\cdot{\bm\nabla}_\br)\,\varrho=[{\sf H},\varrho]\,.
\end{equation}
The ``Hamiltonian matrix'' ${\sf H}$ depends on the phase-space variables
$(t,\br,E,\bv)$. It is explicitly
\begin{equation}
{\sf H}=\frac{{\sf M}^2}{2E}+\sqrt2\GF
\[{\sf N}_\ell+\int d\Gamma'\,
(1-\bv\cdot\bv')\,\varrho_{t,\br,E',\bv'}\],
\end{equation}
where ${\sf M}^2$, the matrix of neutrino mass-squares, is what drives vacuum
oscillations. The matrix of charged-lepton densities, ${\sf N}_\ell$,
provides the usual Wolfenstein matter effect, assuming the background medium
is isotropic. The neutrino and antineutrino phase-space integration $\int
d\Gamma'$ is explicitly \smash{$\int_{-\infty}^{+\infty}dE'E'^2\int
d{\bv}'/(2\pi)^3$}, where the velocity integration is over the unit sphere.
The factor $(1-\bv\cdot\bv')$ represents the current-current nature of
low-energy weak interactions. As a consequence, collinear neutrino modes do
not influence each other and neutrino-neutrino refraction involves direction
of motion (``multi angle effects'') as a central ingredient.

Early studies of neutrino-neutrino refraction were motivated by cosmology and
considered a homogeneous system evolving as a function of time. All
directional effects in Eq.~(\ref{eq:EOM1}) were integrated out. The matter
effect also dropped out, being common to all modes, leading to a much simpler
equation. A homogeneous system evolving in time remains an important
theoretical laboratory to develop an understanding of collective flavor
evolution. We will use it later in its incarnation of a ``colliding neutrino
beam,'' reducing it to one spatial dimension.

\begin{figure}
\centering
\includegraphics[width=0.8\textwidth]{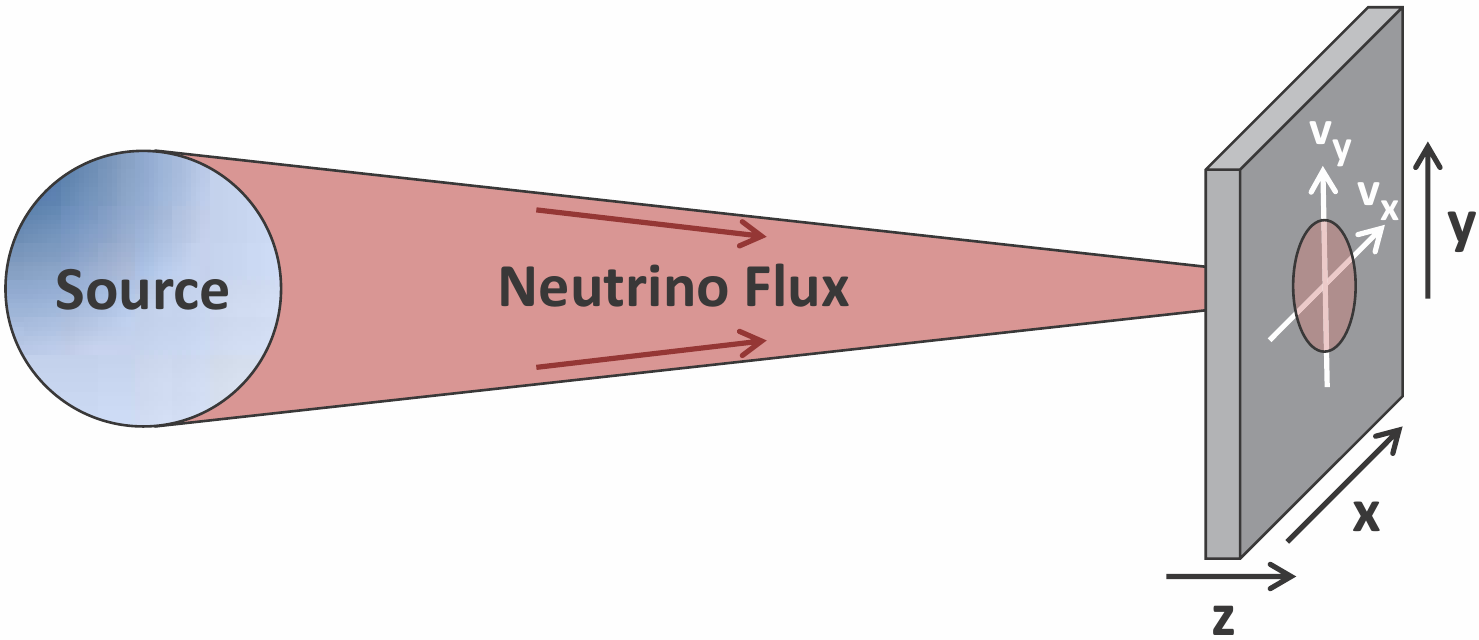}
\caption{Geometry for flavor evolution of neutrinos streaming from
a compact source. The current-current nature of weak interactions
implies that neutrino-neutrino refraction
depends on relative velocities in the transverse plane
($x$--$y$--plane) caused by the finite size of the source.
In a linearized stability analysis we look for modes which grow
exponentially in flavor space as a function of~$z$.} \label{fig:geometry}
\end{figure}

However, the main interest in this subject is provided by core-collapse
supernovae or neutron-star mergers where neutrinos stream away from a compact
source and we ask for flavor evolution as a function of distance. The
current-current nature of weak interactions implies that collinear neutrinos
do not affect each other. Therefore, neutrino-neutrino refraction depends on
the finite size of the source which implies that at some distance, the local
neutrino field has transverse velocity components corresponding to their
origin at different source locations as shown in Fig.~\ref{fig:geometry}. If
the source is spherical with radius $R$, the transverse velocities at the
test distance $r$ fill a circle with radius (in velocity space) of
$(v_x^2+v_y^2)^{1/2}\leq r/R$ as shown in Fig.~\ref{fig:geometry}. The same
applies to a non-spherical source if it is compact and we can think of $R$ as
an envelope.

When we perform a linearized stability analysis we consider a small volume at
some distance $r$ from the source and search for modes of the neutrino mean
field with off-diagonal elements that grow exponentially as a function of
distance. We identify the latter with the $z$-direction, whereas $x$ and $y$
are cartesian coordinates in the transverse plane. Notice that this picture
also applies to a non-spherical source geometry, where one would identify the
local $z$-direction with the neutrino flux direction \cite{Dasgupta:2008cu}.
In this case, of course, the neutrino field in the transverse plane is a
nontrivial function of $x$ and $y$ and of the transverse velocity components
$v_x$ and $v_y$. In the following, however, we will always consider a
spherical source.

This or similar geometric arrangements have been used in all studies of
supernova neutrino flavor evolution, yet may not be adequate to capture
crucial aspects of the problem. One issue arises because neutrinos
occasionally scatter on their way out, providing a small ``halo flux.'' In
other words, the transverse velocities of the ``neutrino beam'' do not only
form a compact disk as shown in Fig.~\ref{fig:geometry}, but in addition have
a diffuse halo \cite{Cherry:2012zw, Sarikas:2012vb}. Even though this halo is
orders of magnitude smaller than the primary beam, it has large angular
leverage and thus a strong refractive impact. Notice that neutrinos in the
main beam are nearly collinear so that $(1-{\bf v}\cdot{\bf v}')$ suppresses
the effective interaction energy by an approximate factor $(R/r)^2$ with
distance, in addition to the geometric $r^{-2}$ flux dilution, i.e.,
$\mu_{\rm eff}\propto r^{-4}$ for the beam component \cite{Duan:2006an}.

Most studies have modeled the neutrino flux as a ``compact beam,'' but
including broad tails in the transverse velocity distribution is a
quantitative issue, not a conceptual one. However, the halo flux extends to
all directions so that at the test location, there is also a ``backward'' or
``inbound'' neutrino flux which has never been included. If neutrinos move in
all directions it is not obvious that we can describe the flavor evolution as
a function of distance. In a linearized stability analysis, it is not obvious
that we should look only for modes which grow exponentially with distance
from the source. This issue is exacerbated if flavor instabilities occur in
the decoupling or ``neutrino sphere'' region as has been speculated by
several authors \cite{Sawyer:2005jk, Duan:2014gfa, Mirizzi:2015fva,
Dasgupta:2015iia, Sawyer:2015dsa}. One may wonder if in this case we should
think of flavor evolution as a collective phenomenon as a function of a
single space or time variable. More plausibly, it represents a nontrivial
space-time phenomenon, but has never been treated as such in the literature.

One may speculate that a linearized stability analysis remains useful in the
sense of a self-consistency test. If one finds stability, self-induced flavor
conversion likely do not occur. On the other hand, if instability is found in
regions where neutrinos stream in all directions, what happens realistically
is a matter of speculation at present. In this spirit we limit ourselves to
the method of linearized stability analysis for the rest of our discussion.

\section{Linearized equations of motion}
\label{sec:linearized}

As a first simplification we limit ourselves to two flavors $\nu_e$ and
$\nu_x$, where $\nu_x$ is a suitable mixture of $\nu_\mu$ and $\nu_\tau$. We
use the unit vectors in flavor space $\vec{B}$ pointing in the mass direction
and $\vec{L}$ in the weak-interaction direction.\footnote{We denote vectors
in flavor space with an arrow, vectors in coordinate and momentum space in
boldface.} Their relative angle is $2\Theta$, twice the mixing angle, which
is taken to be small. We also use the vacuum oscillation frequency
$\omega=\Delta m^2/2E$, where $\Delta m^2$ is a positive number. Henceforth
we label neutrino modes by $\omega$ instead of $E$, where positive $\omega$
is for neutrinos and negative $\omega$ for antineutrinos. Flavor oscillations
are driven by
\begin{equation}\label{eq:ham1}
{\sf H}=\frac{1}{2}\(\omega\vec{B}+\lambda\vec{L}\)\cdot\vec{\sigma}
+\sqrt2\GF\int d\Gamma'\, (1-\bv\cdot\bv')\,\varrho_{t,\br,\omega',\bv'}\,.
\end{equation}
It is understood that a Jacobian for the $E\to\omega$ variable transformation
has been included in the definition of $\varrho_{t,\br,\omega,\bv}$ which is
a phase-space density in the new variables.  We use $\lambda=\sqrt{2}\GF n_e$
with $n_e$ the net electron density (electrons minus positrons) and
$\vec{\sigma}$ is a vector of Pauli matrices. The matter effect is always
large ($\lambda\gg\omega$) and we can go to a frame rotating with frequency
$\lambda$ around the $\vec{L}$ direction. In this frame, $\vec{B}$ rotates
fast and we assume its components transverse to $\vec{L}$ average to zero.
Equivalently, we can take the mixing angle in matter to be infinitely small.
Either way, in effect $\vec{B}$ and $\vec{L}$ are taken to be collinear.

To determine stability against self-induced flavor conversion we study the
evolution of small deviations from a system initially prepared in flavor
eigenstates. To this end we express the mean-field matrices in terms of
occupation numbers in the form
\begin{equation}
\varrho=\frac{f_{\nu_e}+f_{\nu_x}}{2}\,\mathbb{1}+\frac{f_{\nu_e}-f_{\nu_x}}{2}\,
\begin{pmatrix}s&S\\S^*&-s\end{pmatrix}\,,
\end{equation}
where all quantities depend on the phase-space variables
$(t,\br,\omega,\bv)$. For antineutrinos, the coefficients are
$-f_{\nu_e}-f_{\nu_x}$ and $-f_{\nu_e}+f_{\nu_x}$, respectively. The first
term is not affected by flavor conversion. The matrix structure is encoded in
the real number $s$ and the complex number $S$ with $s^2+|S|^2=1$.
Equivalently, the matrix structure can be expressed in terms of a normalized
polarization vector $\vec{P}$ as $\vec{P}\cdot\vec{\sigma}$, corresponding to
$S=P_1+\I P_2$ and $s=P_3$. We will linearize in $S$, the off-diagonal
element of $\varrho$, and note that to this order $s=1$.

As a further simplification we restrict ourselves to a neutrino gas with
constant properties over the distances and time scales of flavor conversion,
i.e., we assume that the initial occupation numbers $f$ depend only on
$\omega$ and $\bv$. The linearized equation of motion ($|S|\ll1$) is
\begin{equation}\label{eq:EOM2}
\I(\partial_t+\bv\cdot{\bm\nabla}_\br)\,S_{t,\br,\omega,\bv}=
(\omega+\lambda)\,S_{t,\br,\omega,\bv}
+\sqrt{2}\GF\int
d\Gamma'\,(1-\bv\cdot\bv')\,h_{\omega',\bv'}\(S_{t,\br,\omega,\bv}-S_{t,\br,\omega',\bv'}\)\,,
\end{equation}
where $h=f_{\nu_e}-f_{\nu_x}$ for $\omega>0$ and
$h=-(f_{\bar\nu_e}-f_{\bar\nu_x})$ for $\omega<0$. Next we introduce an
effective neutrino density
\smash{$n_\nu=\half(n_{\nu_e}+n_{\bar\nu_e}-n_{\nu_x}-n_{\bar\nu_x})$} and
the neutrino spectrum
\begin{equation}
g_{\omega,\bv}=\frac{h_{\omega,\bv}}{n_\nu}=\frac{1}{n_\nu}\times
\begin{cases}f_{\nu_e}(\omega,\bv)-f_{\nu_x}(\omega,\bv)&\hbox{for $\omega>0$,}\\
-f_{\bar\nu_e}(\omega,\bv)+f_{\bar\nu_x}(\omega,\bv)&\hbox{for $\omega<0$}.\end{cases}
\end{equation}
Our definitions imply the normalization $\int d\Gamma\,{\rm
sign}(\omega)\,g_{\omega,\bv}=2$. We also introduce the parameters
$\epsilon=\int d\Gamma\,g_{\omega,\bv}$ and ${\bm \phi}=\int
d\Gamma\,\bv\,g_{\omega,\bv}$ which express the asymmetry of the $\nu$ vs.\
$\bar\nu$ density and flux, respectively. In terms of these quantities, the
linearized equation becomes
\begin{equation}\label{eq:EOM3}
\I(\partial_t+\bv\cdot{\bm\nabla}_\br)\,S_{t,\br,\omega,\bv}=
\Bigl[\omega+\lambda+\mu\,(\epsilon-\bv\cdot{\bm\phi})\Bigr]\,S_{t,\br,\omega,\bv}-
\mu \int d\Gamma'\,(1-\bv\cdot\bv')\,g_{\omega',\bv'}S_{t,\br,\omega',\bv'}\,,
\end{equation}
where the effective neutrino-neutrino interaction energy is $\mu=\sqrt{2}\GF
n_\nu$. The term $\mu\,(\epsilon-\bv\cdot{\bm\phi})$ is the ordinary matter
effect caused by neutrinos on each other irrespective of collective effects.
The matter background quantified by $\lambda$ also provides a flux term if it
is not isotropic.

Our equations are formulated for inverted neutrino mass ordering in our
two-flavor system. Normal ordering corresponds to $\vec{B}\to -\vec{B}$ in
Eq.~(\ref{eq:ham1}) and thus to a minus sign in front of the $\omega$ term in
square brackets in Eq.~(\ref{eq:EOM3}). In a stability analysis, one can
instead flip the sign of $\mu$ and $\lambda$ and extend these parameters to
the range $-\infty<\mu,\lambda<+\infty$ to cover both cases of mass ordering.

On the level of a linearized stability analysis, all examples studied in the
literature are special cases of Eq.~(\ref{eq:EOM3}) with different
assumptions about the neutrino energy and velocity distribution, about the
initial or boundary conditions, and about the symmetries of the solutions.

\section{Spontaneous symmetry breaking in one dimension}
\label{sec:beam}

\subsection{Colliding neutrino beam}

The main conceptual point of our discussion is spontaneous symmetry breaking
in collective flavor conversion, i.e., the solution does not respect the
symmetry of the initial or boundary condition, leading to entirely new
solutions than had been conceived of earlier. Assuming a stationary spherical
source which emits neutrinos isotropically, the local neutrino field at our
test position as in Fig.~\ref{fig:geometry} will be axially symmetric
(isotropic in the transverse plane). The solution can be axially symmetric,
but can also break axial symmetry, which was termed the multi-azimuth angle
(MAA) solution \cite{Raffelt:2013rqa}. While the former solution appears for
inverted neutrino mass ordering, the latter requires normal ordering. The
symmetry-breaking solution varies as $\cos\varphi$ or $\sin\varphi$ in terms
of the azimuth angle $\varphi$ in the $x$--$y$--plane, i.e., it has dipole
structure, essentially as a consequence of the current-current nature of the
interaction term.

In the geometry of Fig.~\ref{fig:geometry}, the crucial effects occur in the
transverse plane, with distance $z$ playing the role of a parameter that
describes the evolution. Assuming a stationary source and solution, time
drops out entirely and we may rename $z$ as ``time'', an affine parameter
along the radial direction. In this sense, the stationary radial evolution of
supernova neutrinos has often been described in terms of ``time evolution.''
The neutrino density decreases as a function of distance and transverse
scales increase, so one can visualize the evolution of the stationary
neutrino field as a function of distance instead as a time evolution in an
expanding space, analogous to the evolution in the expanding universe
\cite{Chakraborty:2015tfa}. In this sense the two cases bear many
similarities.

\begin{figure}
\centering
\includegraphics[width=0.40\textwidth]{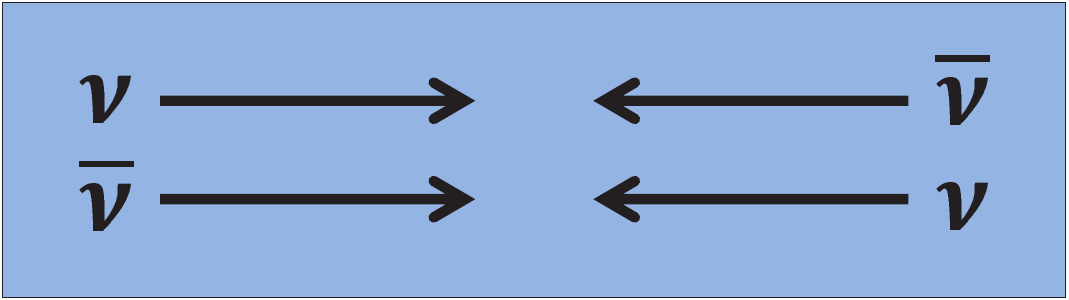}
\caption{Initially homogeneous ensemble of four neutrino modes (``colliding beams''
of neutrinos and antineutrinos). The system is taken to be infinite in all
directions. The normalized $\nu$ flux is $1+a$, the $\bar\nu$ flux
$1-a$ with the asymmetry parameter $a$ in the range
$-1\leq a\leq+1$. The left-right asymmetry is parametrized by $b$ such that
the upper beam in this figure has normalized strength $1+b$, the lower beam
$1-b$ with $-1\leq b\leq+1$.}\label{fig:beam}
\end{figure}

Axial symmetry breaking of supernova neutrinos leads to a dipole variation in
the transverse plane, which on the crudest level of approximation can be seen
as two azimuth bins, corresponding to a beam of left-moving and a beam of
right-moving neutrinos (in the transverse plane). This picture suggests to
study simple toy cases of symmetry breaking in terms of a homogeneous system
with two opposite-moving momenta and study it as a function of time
\cite{Raffelt:2013isa} in the spirit of Fig.~\ref{fig:beam}. One may consider
this abstract system as a colliding neutrino beam, initially homogeneous and
evolving in time, or as representing the transverse behavior of neutrinos
streaming from a compact source. Other authors have constructed an analogous
system as the ``line model'' of neutrino emission \cite{Duan:2014gfa,
Abbar:2015mca}. It has also been called a ``two-dimensional model''
\cite{Mirizzi:2015fva}, where one dimension is our beam, the other the
$z$-direction which we here call ``time.'' Different authors prefer different
visualizations of the same physical content.

We formulate this problem in terms of the colliding beam model of
Fig.~\ref{fig:beam}, where the possible velocities are $v=\pm1$. (Of course,
the transverse velocities in a geometry like Fig.~\ref{fig:geometry} are very
small, but we can renormalize them by rescaling the interaction
parameter~$\mu$.) In addition we use monochromatic neutrinos and
antineutrinos with energies $\pm E$ and thus with vacuum oscillation
frequencies $\pm\omega=\pm\Delta m^2/2E$. Therefore, we represent the
function $S_{t,x,\omega,v}$ as a quartet of numbers $(R,\bar L,L,\bar R)$,
representing the off-diagonal elements of the left- and right-moving
neutrinos and antineutrinos, respectively, which all depend on $t$ and one
spatial coordinate $x$. Moreover, we denote the phase-space densities
$g_{\omega,v}$ for our four modes as $(r,\bar l,l,\bar r)$. The linearized
equations of motion can be transformed to Fourier space by considering
solutions of the form $S_{t,\br}=Q_{\Omega,\bk} e^{-\I(\Omega
t-\bk\cdot\br)}$, leading to
\begin{equation}\label{eq:EOM4}
\Omega\,\begin{pmatrix}R\\ \bar L\\ L\\ \bar R\end{pmatrix}=\left[
\begin{pmatrix}\omega+k&0&0&0\\0&-\omega-k&0&0\\0&0&\omega-k&0\\0&0&0&-\omega+k\end{pmatrix}
+\lambda+2\mu
\begin{pmatrix}
l+\bar l&-\bar l&-l&0\\
-r&r+\bar r&0&-\bar r\\
-r&0&r+\bar r&-\bar r\\
0&-\bar l&-l&l+\bar l\\
\end{pmatrix}
\right]\begin{pmatrix}R\\ \bar L\\ L\\ \bar R\end{pmatrix}\,.
\end{equation}
The factor of 2 in the neutrino-neutrino interaction term arises from the
factor $(1-\bv\cdot\bv')$ being 2 for opposite moving modes and 0 for
parallel ones.

This is the most general one-dimensional four-mode case, which we simplify
somewhat further. We assume the overall normalized neutrino density to be
$1+a$, the antineutrino density $1-a$ so that $-1<a<+1$ quantifies the
$\nu$--$\bar\nu$--asymmetry. Moreover, we picture the system as consisting of
the beam $R$ and $\bar L$ (upper modes in Fig.~\ref{fig:beam}) with overall
strength $1+b$, and the beam $L$ and $\bar R$ (lower modes in
Fig.~\ref{fig:beam}) with abundance $1-b$ so that $-1<b<+1$ describes an
initial left-right asymmetry. Therefore, the phase-space densities of our
four modes are taken to be
\begin{subequations}\label{eq:beamoccupations}
\begin{eqnarray}
r     &=& +\half(1+a)(1+b)\,,\\
\bar l&=& -\half(1-a)(1+b)\,,\\
\bar r&=& -\half(1-a)(1-b)\,,\\
l     &=& +\half(1+a)(1-b)\,.
\end{eqnarray}
\end{subequations}
The overall normalization is indeed $r+l-\bar r-\bar l=2$ for all $a$ and
$b$. The neutrino-neutrino interaction matrix in Eq.~(\ref{eq:EOM4}) becomes
\begin{equation}\label{eq:abmatrix}
\mu\begin{pmatrix}
2\,(a-b)&(1-a)(1+b)&~-(1+a)(1-b)&0\\
-(1+a)(1+b)&2\,(a+b)&~0&(1-a)(1-b)\\
-(1+a)(1+b)&0&~2\,(a+b)&(1-a)(1-b)\\
0&(1-a)(1+b)&~-(1+a)(1-b)&2\,(a-b)\\
\end{pmatrix}\,.
\end{equation}
Overall we have to deal with a quartic eigenvalue equation.

\subsection{Spontaneous left-right symmetry breaking}

As a first explicit case we begin with a system which is prepared to be
left-right symmetric ($b=0$). If the solution inherits this symmetry it
should be left-right symmetric as well, so it makes sense to define
left-right symmetric modes $A_+=\half(R+L)$ and $\bar A_+=\half(\bar R+\bar
L)$ as well as antisymmetric modes $A_-=\half(R-L)$ and $\bar A_+=\half(\bar
R+\bar L)$. The eigenvalue equation then is
\begin{equation}\label{eq:EOM5}
\Omega\,\begin{pmatrix}A_+\\ \bar A_+\\ A_-\\ \bar A_-\end{pmatrix}=
\left[
\begin{pmatrix}\omega&0&k&0\\0&-\omega&0&k\\k&0&\omega&0\\0&k&0&-\omega\end{pmatrix}
+\lambda+2a\mu+\mu
\begin{pmatrix}
-(1+a)&1-a&0&0\\
-(1+a)&1-a&0&0\\
0&0&1+a&-(1-a)\\
0&0&1+a&-(1-a)\\
\end{pmatrix}
\right]
\begin{pmatrix}A_+\\ \bar A_+\\ A_-\\ \bar A_-\end{pmatrix}\,.
\end{equation}
Assuming the solution to be homogeneous ($k=0$), the left-right symmetric and
antisymmetric modes decouple and we find two pairs of equations.

The upper pair (symmetric solution) corresponds to the traditional flavor
pendulum~\cite{Duan:2005cp, Hannestad:2006nj}. The eigenvalues are explicitly
\begin{equation}
\Omega=\lambda+a\mu\pm\sqrt{(a\mu)^2+\omega(\omega-2\mu)}\,.
\end{equation}
The argument of the square root is negative, and thus we have a run-away
solution, in the range
\begin{equation}
\frac{1}{1+\sqrt{1-a^2}}<\frac{\mu}{\omega}<\frac{1}{1-\sqrt{1-a^2}}\,.
\end{equation}
The maximum growth rate arises for $\mu=\omega/a^2$ and is ${\rm
Im}\,\Omega|_{\rm max}=\omega\sqrt{1-1/a^2}$. For normal mass ordering, we
should take $\omega\to-\omega$ or equivalently $\mu\to-\mu$ as explained
earlier. In this case $\Omega$ is always real and one finds no instability.
This basic case illustrates what seemed to be a generic property of
self-induced flavor conversion: there is an instability for inverted mass
ordering and the growth rate is of order $\Delta m^2/2E$.

The lower pair of equations in Eq.~(\ref{eq:EOM5}) is the same as the upper
one with $\mu\to-\mu$. Therefore, the left-right antisymmetric modes have the
same instability as the symmetric case, but for normal neutrino mass
ordering. This observation was the main point of Ref.~\cite{Raffelt:2013isa},
where a graphical explanation in terms of polarization vectors was given. The
main lesson is that unstable solutions do not need to respect the symmetries
of the initial condition, but of course the unstable mode requires a small
seed (a small disturbance of the symmetric initial condition) to be able to
grow.

\subsection{Spontaneous breaking of homogeneity}

We may next consider the evolution of a spatial Fourier mode with
nonvanishing wave number $k$, keeping the system initially left-right
symmetric ($b=0$). The idea to consider the evolution and stability of
spatial Fourier modes was first proposed rather
recently~\cite{Mangano:2014zda, Duan:2014gfa} and more detailed studies
followed soon \cite{Mirizzi:2015fva, Abbar:2015mca, Chakraborty:2015tfa}. In
our simple beam case, the left-right symmetric and antisymmetric modes are
coupled so that it is simpler to return to the original version
Eq.~(\ref{eq:EOM4}) together with Eq.~(\ref{eq:abmatrix}). With $b=0$, the
eigenvalue equation is
\begin{equation}\label{eq:ev1}
{\rm det}\[\lambda+2a\mu-\Omega+
\begin{pmatrix}
\omega+k&(1-a)\mu&-(1+a)\mu&0\\
-(1+a)\mu&-\omega-k&0&(1-a)\mu\\
-(1+a)\mu&0&\omega-k&(1-a)\mu\\
0&(1-a)\mu&-(1+a)\mu&-\omega+k&\\
\end{pmatrix}\]=0\,.
\end{equation}
The explicit solutions of this quartic equation are too complicated to be
informative. However, in the limit $k\to\pm\infty$ it can be simplified. In
this limit, the diagonal elements of the matrix strongly dominate and we
expect the solution to be of the form $\Omega=\tilde\Omega+\lambda+2a\mu\pm
k$, where $\tilde\Omega$ is a frequency which is small compared with $k$. We
first consider $\Omega=\tilde\Omega+\lambda+2a\mu-k$, leading to
\begin{equation}
{\rm det}
\begin{pmatrix}
2k+\omega-\tilde\Omega&(1-a)\mu&-(1+a)\mu&0\\
-(1+a)\mu&-\omega-\tilde\Omega&0&(1-a)\mu\\
-(1+a)\mu&0&\omega-\tilde\Omega&(1-a)\mu\\
0&(1-a)\mu&-(1+a)\mu&2k-\omega-\tilde\Omega&\\
\end{pmatrix}=0\,.
\end{equation}
In the limit $k\to\infty$ we may approximate the upper left and lower right
diagonal elements with $2k$, leading to a quadratic equation for the
eigenvalues $\tilde\Omega$. The expressions simplify considerably if we write
$\mu=m\sqrt{\omega k}$ in terms of a dimensionless parameter $m$, leading to
\begin{equation}
\tilde\Omega=\(-a^2m^2\pm\sqrt{1-2a m^2+a^4 m^4}\)\omega\,.
\end{equation}
For this expression to have an imaginary part, we need $a>0$, corresponding
to an excess of neutrinos over antineutrinos, and in a range of $m$ values
corresponding to
\begin{equation}
a\(1-\sqrt{1-a^2}\)<\frac{k\omega}{\mu^2}<a\(1+\sqrt{1-a^2}\)\,.
\end{equation}
For $k\to-\infty$ and $a>0$ the system is stable. Unstable solutions also
exist when both $a$ and $k$ are negative, i.e., the system has unstable modes
if $a$ and $k$ have equal sign. The unstable $\mu$ range scales with
$\sqrt{\omega k}$, in agreement with previous findings \cite{Duan:2014gfa,
Chakraborty:2015tfa}.

Notice that the sign of $k$ distinguishes between left- or right-moving
spatial disturbances, i.e., the spontaneous breaking of homogeneity by the
exponential growth of $k$ modes also breaks the initial left-right symmetry
of the system.

The Fourier mode with the maximum growth rate is $k=a^3\mu^2/\omega$ and
provides ${\rm Im}\,\Omega|_{\rm max}=\omega\,(a^{-2}-1)^{1/2}$. For any $k$
and taking $\omega=0$, the eigenvalue equation~(\ref{eq:ev1}) has only real
solutions. The speed of growth of the unstable modes in this system is always
of order $\omega$.

Going beyond linear order, the equations of motion couple different spatial
Fourier modes because in Fourier space, the rhs of Eq.~(\ref{eq:EOM1})
becomes a convolution. Therefore, spatial inhomogeneities will quickly lead
to kinematical decoherence \cite{Mirizzi:2015fva}. One may think that this
outcome is inevitable because for any neutrino density (any value of $\mu$)
there is a range of unstable $k$-modes. On the other hand, in the geometry of
Fig.~\ref{fig:geometry} we also need to include the multi-angle matter effect
and study the instability condition in the parameter space spanned by $\mu$,
$\lambda$ and $k$. In this geometry, the meaning of ${\bf k}$ is a Fourier
mode in the transverse plane and the spontaneous breaking of homogeneity
actually amounts to the breaking of global spherical symmetry. As it has been
seen several times \cite{EstebanPretel:2008ni, Sarikas:2011am,
Sarikas:2012vb, Chakraborty:2011nf, Chakraborty:2011gd}, the inclusion of
multi-angle matter effects tends to suppress instabilities when the matter
potential is large. For the inhomogeneous instabilities this also turns out
to be true \cite{Chakraborty:2015tfa} (see also
Fig.~\ref{fig:inhomogeneous}), and the unstable region of parameter space,
the instability ``footprint'', is pushed to larger values of $\mu$ for a
large $\lambda$. Another interesting consequence of the multi-angle matter
effect seems to be that the largest spatial scales (small $k$) in supernovae
are the most unstable ones. This, in the sense that larger values of $k$ give
rise to instability footprints that are further away from the density profile
of a realistic supernova than for $k=0$, meaning that all transverse $k$
modes are stable if the homogeneous mode is stable.  On the other hand, if an
instability is encountered, flavor decoherence may be a generic outcome.

\section{Fast flavor conversion}
\label{sec:fastflavor}

\subsection{Asymmetric beam}

So far the growth rate of unstable modes was found to be proportional
to the vacuum oscillation frequency. On the other hand, Sawyer has
shown that one can construct cases where in our notation the growth
rate is proportional to $\mu$, not to $\omega$ \cite{Sawyer:2005jk,
  Sawyer:2015dsa}. Our colliding beam model provides perhaps the
simplest possible example for this phenomenon of ``fast flavor conversion,''
but it requires explicit left-right
symmetry breaking by the initial condition \cite{Chakraborty:2016lct}.
Setting $b=1$ in
Eq.~(\ref{eq:beamoccupations}) implies that in Fig.~\ref{fig:beam}
only the upper modes are occupied, the lower ones are empty: we have only
right-moving $\nu$ and only left-moving $\bar\nu$. The equation of
motion for the two occupied modes following from Eqs.~(\ref{eq:EOM4})
and (\ref{eq:abmatrix}) is
\begin{equation}\label{eq:EOM6}
\Omega\,\begin{pmatrix}R\\ \bar L\end{pmatrix}
=\left[\begin{pmatrix}\omega+k&0\\0&-\omega-k\end{pmatrix}+2\mu
\begin{pmatrix}-1+a,&1-a\\-1-a,&1+a\end{pmatrix}
\right]\begin{pmatrix}R\\ \bar L\end{pmatrix}\,.
\end{equation}
We notice immediately that the role $\omega$ is here played by
$\tilde\omega=\omega+k$. Even for vanishing $\omega=\Delta m^2/2E$, we
find unstable solutions in the form of spatial Fourier modes with
$k\not=0$. The reason for this behavior is
that neutrinos (vacuum frequency $+\omega$) move
right so that the spatial Fourier term ${\bf v}\cdot{\bf k}$ enters as $+k$,
and the other way round for the beam of left-moving antineutrinos.

The explicit eigenvalues are found to be
$\Omega=2a\mu\pm[(2a\mu)^2+\tilde\omega(\tilde\omega-4\mu)]^{1/2}$,
i.e., for $\tilde\omega=0$ they are purely real. One finds an
imaginary part for
$1-(1-a^2)^{1/2}<\tilde\omega/2\mu<1+(1-a^2)^{1/2}$.
Because $\mu$ is defined to be positive, we have unstable
solutions only for $\tilde\omega>0$ or $-\omega<k$.
For a fixed $\mu$ value, the maximum growth rate occurs for
$\tilde\omega=2\mu$ and is
\begin{equation}
{\rm Im}\,\Omega\vert_{\rm max}
=2\mu\sqrt{1-a^2}\,.
\end{equation}
This rate is indeed ``fast'' in the sense that it is proportional to
$\mu=\sqrt{2}\GF n_\nu$.

The explicit left-right symmetry breaking of the initial system need
not be maximal to obtain this effect. The stability in the entire
parameter space $-1<a<1$ and $-1<b<1$ was studied in
Ref.~\cite{Chakraborty:2016lct}.  A certain range of unstable $k$ modes exists
if $a^2<b^2$, i.e., the left-right asymmetry must exceed the
neutrino-antineutrino asymmetry.  Details aside, the main point is
that a sufficiently asymmetric system can be unstable in flavor space
even if neutrinos were exactly massless and even if vacuum flavor
oscillations would not exist.

\subsection{Different zenith angle distributions for neutrinos and antineutrinos}

In the supernova geometry of Fig.~\ref{fig:geometry}, the ``left-right
asymmetry'' of the previous section would translate into significant
azimuthal asymmetries of the neutrino and antineutrino velocities in
the transverse plane.  However, fast flavor conversion can also appear
in a different case of explicit symmetry breaking, i.e., when the
$\nu_e$ and $\bar\nu_e$ zenith angle distributions are sufficiently
different \cite{Sawyer:2015dsa}.  These species have rather different
interaction rates with a nuclear medium and thus decouple in different
regions. Typically, one expects a larger flux of $\nu_e$ because of
deleptonization, and a broader zenith-angle distribution for $\nu_e$
because of their larger interaction rate. However, in different phases
or regions of a core-collapse supernova or in neutron-star mergers, a
variety of distributions may occur. We here worry only about the
conceptual issue of fast flavor conversion, not about realistic
quantitative scenarios.

This system can be mimicked with a small variation of our colliding beam
model. The zenith-angle distribution in the geometry of
Fig.~\ref{fig:geometry} translates into a distribution of velocities in the
transverse plane. Neutrinos emitted in the radial direction have vanishing
velocity in the observation plane, whereas those emitted from the limb of the
source have a maximal transverse velocity of $v_{\rm max}=R/r$ (source radius
$R$, distance $r$). Therefore, the zenith-angle distribution can be
represented by a colliding beam with different velocities. We can use any
convenient range of velocities, absorbing an overall scale in the definition
of the interaction strength $\mu$. So we consider a colliding beam as in
Fig.~\ref{fig:beam}, the crucial ingredient being a different velocity for
$\nu_e$ and $\bar\nu_e$, while maintaining perfect left-right symmetry. This
toy example mimics the idea that $\nu_e$ and $\bar\nu_e$ are emitted from
neutrino spheres with different radius, which is the original model studied
by Sawyer \cite{Sawyer:2015dsa}.

Specifically we use $r=l=\half(1+a)$ and $\bar r=\bar l=-\half(1-a)$ for the
occupations of the neutrino and antineutrino modes with the same asymmetry
parameter $-1<a<1$ as before. Moreover, we use the velocities $v_R=-v_L=1+b$
and $v_{\bar R}=-v_{\bar L}=1-b$ so that $-1<b<1$ expresses the different
``zenith angles.'' The left-right symmetric setup and using the homogeneous
case $k=0$ implies that the left-right symmetric and antisymmetric modes
decouple. In the limit $\omega=0$, the eigenvalues for the latter are purely
real and thus do not show fast flavor conversion. The symmetric modes, on the
other hand, produce the eigenvalues
\begin{equation}
\Omega/\mu=a-2b-ab^2\pm\sqrt{4ab(1+b^2)+a^2(1+6b^2+b^4)}\,.
\end{equation}
A necessary condition for a nonvanishing imaginary part is $ab<0$, meaning
that the species $\nu_e$ or $\bar\nu_e$ with the larger abundance must have
the smaller ``zenith angle'' for a fast instability to exist. The main point
is that we can easily construct a homogeneous and left-right symmetric beam
model that shows fast flavor conversion.

More physical examples in the spirit of Fig.~\ref{fig:geometry} were studied
in Refs.~\cite{Sawyer:2015dsa, Chakraborty:2016lct}, assuming neutrino ``bulbs'' with
different radii for $\nu_e$ and $\bar\nu_e$. The exact conditions where fast
flavor conversion appears~\cite{Chakraborty:2016lct} depends on the matter effect
$\lambda$ besides the neutrino zenith-angle distributions. The appearance of
this effect does not seem to require extreme parameters.

\section{Temporal Instabilities}
\label{sec:temporal}

In principle it is straightforward to solve the general equation of motion
Eq.~(\ref{eq:EOM1}) or its linearized version Eq.~(\ref{eq:EOM3}) in Fourier
space, although it took a long time until it dawned on our community that the
nature of self-induced flavor conversion as an instability implies that, for
example, an initially homogeneous system can grow large inhomogeneities in
flavor space if some spatial Fourier modes are unstable. A similar
observation applies to a Fourier transform in the time domain. We can imagine
neutrinos streaming from a stationary source and we may ask for the flavor
evolution as a function of distance, for the moment assuming perfect
spherical symmetry of the source and the solution.

As a first case we may imagine that the source has some small periodic time
variation with frequency $\Omega$ imprinted on it \cite{Abbar:2015fwa}. Then,
instead of solving the stationary version of Eq.~(\ref{eq:EOM1}) without time
variation of the source or solution, we should instead consider a mode
varying as $e^{-\I\Omega t}$. In particular in the linearized two-flavor form
of Eq.~(\ref{eq:EOM3}), the only consequence is that
$\lambda\to\lambda-\Omega$. As a next step, we should proceed as before and
find, for example, growing modes as a function of radial distance. This
approach was taken one step further in Ref.~\cite{Dasgupta:2015iia} where the
authors argued that we do not need a driving frequency $\Omega$ provided by
some supernova physics, but that a small temporal disturbance was enough to
trigger a growing mode of that frequency.

The crucial physical impact of the modification $\lambda\to\lambda-\Omega$ is
that the matter effect is reduced. As we already discussed, the presence of
$\lambda$ has a strong impact on the flavor evolution as a function of
distance through the multi-angle matter effect \cite{EstebanPretel:2008ni,
Sarikas:2011am, Sarikas:2012vb, Chakraborty:2011nf, Chakraborty:2011gd}. As a
simple case we consider plane-parallel geometry and study the evolution as a
function of $z$, assuming translational symmetry in the $x$--$y$--plane. In
this case $\I\bv\cdot{\bm\nabla}$ in Eq.~(\ref{eq:EOM3}) becomes $\I v_z
\partial_z$ and the linearized equation of motion becomes
\begin{equation}\label{eq:EOM8}
\I v_z\partial_z\,S_{\Omega,z,\omega,\bv}=
\Bigl[\omega+\lambda-\Omega+\mu\,(\epsilon-v_z\phi_z)\Bigr]\,S_{\Omega,z,\omega,\bv}-
\mu \int d\Gamma'\,(1-\bv\cdot\bv')\,g_{\omega',\bv'}S_{\Omega,z,\omega',\bv'}\,.
\end{equation}
The impact of $v_z$ appearing on the lhs is most easily understood in the
limit where the neutrino flux is nearly collinear, i.e., at a large distance
from the source. We may describe all modes by the small velocity vector
${\bm\beta}$ in the transverse plane and we bring $v_z$ to the rhs in the
form $v_z^{-1}=(1-{\bm\beta}^2)^{-1/2}\approx1+\half{\bm\beta}^2$. On the rhs
all terms except $\omega+\lambda-\Omega$ are already of order $\beta^2$ and
thus to order ${\bm\beta}^2$ remain unaffected by this factor. The mode
frequencies $\omega$ already vary over a broad range and are small compared
with $\lambda$, so we may approximate
$(1+\half{\bm\beta}^2)\omega\approx\omega$. To solve Eq.~(\ref{eq:EOM8}) we
are looking for solutions proportional to $e^{\I K z}$ with $K$ a complex
wavenumber where an imaginary part of $K$ represents an exponentially growing
mode. Therefore, the eigenvalue equation involves a term
$\omega+(1+\half{\bm\beta}^2)(\lambda-\Omega)+K=
\omega+\half{\bm\beta}^2(\lambda-\Omega)+\tilde K$ with $\tilde
K=K+\lambda-\Omega$. In other words, the real part of $K$ gets shifted,
corresponding to solving the equation in a frame in flavor space which
rotates with frequency $\lambda-\Omega$ as a function of $z$. What remains is
a term $\half{\bm\beta}^2(\lambda-\Omega)$ which is different for every
velocity, i.e., the impact of $\lambda$ depends on direction. This
``multi-angle matter effect''  originates from the simple observation that a
mode with general $\bv$ covers a larger distance to reach the same $z$ than a
mode which moves along the $z$-direction and thus acquires a larger matter
phase.

\begin{figure}
\centering
\includegraphics[width=0.7\textwidth]{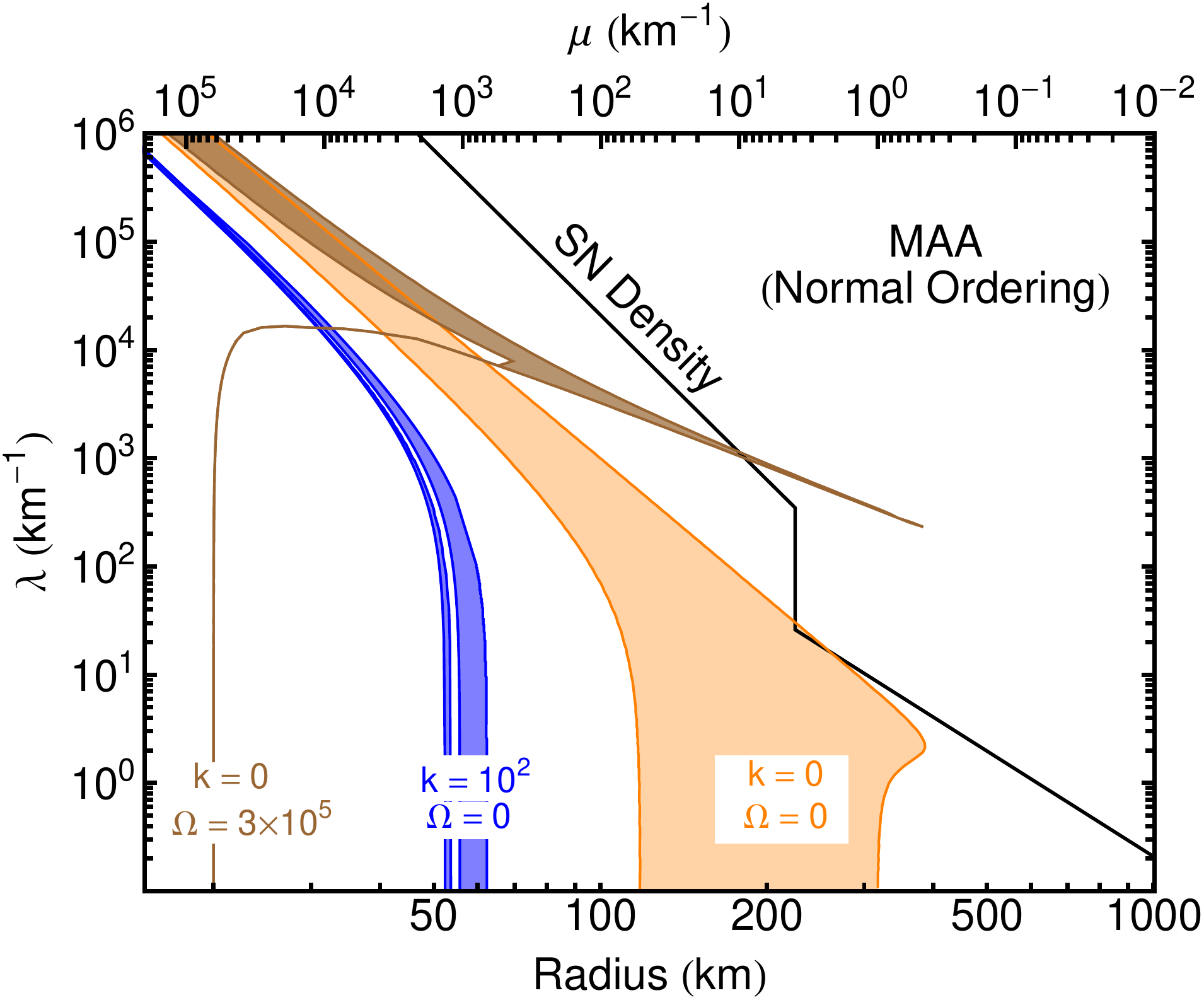}
\caption{Instability footprints of the MAA instability for temporal or spatial
variations for a model with $R=15~\textrm{km}$. The homogeneous and time-independent
instability ($k=0$, $\Omega=0$) is shown for reference. In addition, an
inhomogeneous mode with wave number $k=10^2$ (in units of the vacuum
oscillation frequency)
and a temporal mode with frequency $\Omega=3\times10^5$ are shown. The shaded
areas are defined by $\textrm{Im}\; \Omega > 10^{-2}$. The black line
indicates a schematic density profile of a supernova, where the decrease in
$\lambda$ just above $r=200~\textrm{km}$ corresponds to the shock wave front.
Notice that $\mu \propto r^{-4}$ which gives a one-to-one correspondence
between $\mu$ and radius.} \label{fig:inhomogeneous}
\end{figure}

Neutrinos streaming from a SN core encounter a certain trajectory of
$(\lambda,\mu)$ values and may never encounter any instability if $\lambda$
is large, depending on the ``footprint'' of the unstable region in the
$\lambda$--$\mu$--plane \cite{Sarikas:2011am, Sarikas:2012vb,
Chakraborty:2014lsa, Chakraborty:2015tfa}. In Fig.~\ref{fig:inhomogeneous},
we show such footprints, following the approach from
Ref.~\cite{Chakraborty:2015tfa} for the MAA instability. The footprints
correspond to solutions of Eq.~(\ref{eq:EOM8}) with $\textrm{Im}\; K >
10^{-2}$ for different initial instability seeds. Compared to the notation in
Eq.~(\ref{eq:EOM8}), we have absorbed a factor of $\bm\beta^2$ into the
definitions of $\mu$ and $\lambda$, which means that we redefine
$\lambda=\sqrt{2}G_FN_e(r)(R/r)^2$ and $\mu=\sqrt{2}G_Fn_{\nu_e}(r)(R/r)^2$
where $r$ is the radius and $R$ is the neutrino-sphere radius. The red-orange
footprint is for the stationary, homogeneous mode ($\Omega=0$, $k=0$), and
all the stationary, smaller scale modes fill out the region below this
footprint as demonstrated with the blue footprint for $k=10^2$ which we also
described in Sec.~\ref{sec:beam}. However, with the modification
$\lambda\to\lambda-\Omega(R/r)^2$, at any distance from the source there are
some $\Omega$ modes that are unstable\footnote{The explicit dependence on $R$
in the cancellation condition means that the footprint will depend on the
neutrino-sphere radius. This behavior is in contrast to the stationary case
($\Omega=0$) where the footprints are independent of $R$.}  and grow
exponentially~\cite{Dasgupta:2015iia}. The brown footprint shows the
homogeneous case with a frequency $\Omega=3\times10^{5}$. Here the
cancellation between $\Omega$ and $\lambda$ is obvious as the unstable region
is lifted from small values of $\lambda$ to much larger values. However, it
is also evident that it is only relevant for the supernova profile in a very
narrow distance range. Therefore, the huge growth factors envisioned in
Ref.~\cite{Dasgupta:2015iia} may not necessarily materialize. On the other
hand, one also needs to worry about the impact of spatial Fourier modes in
the transverse direction. Moreover, the question of fast-growing modes for
nontrivial angle distributions as proposed by Sawyer \cite{Sawyer:2015dsa}
have not yet been studied in this context.

\section{Conclusions and Outlook}
\label{sec:conclusions}

The topic of collective neutrino flavor conversion has undergone a shift of
paradigm over the past couple of years. Neutrinos are produced at the source
far from flavor equilibrium due to their different low-energy interaction
channels in stellar core collapse or in neutron-star mergers. Collective
flavor effects seemed to imprint interesting features on the flavor-dependent
spectral fluxes, yet preserve the flavor dependence. However, many of the old
results were based on too many symmetry assumptions about the emission
characteristics and the solutions. Meanwhile it has become clear that the
story of collective neutrino oscillations is a complicated saga of
instabilities in flavor space caused in particular by the spontaneous
breaking of space-time symmetries. Moreover, depending on the angle
distributions, collective flavor conversion does not even depend on neutrino
masses and mixing parameters, a phenomenon dubbed ``fast flavor conversion''
because the exponential growth rate is of order the neutrino interaction
energy $\mu$ instead of the much smaller vacuum oscillation frequency
$\omega$. Several authors have speculated that these different instabilities
and their nonlinear coupling with each other might push the system close to
kinematical decoherence in flavor space. In this sense, the final outcome
might be simple.

However, this development poses new questions. If self-induced flavor
conversion and concomitant decoherence already commence in the neutrino
decoupling region, the traditional treatment in the spirit of
Fig.~\ref{fig:geometry} would not be appropriate. One would have to deal with
an environment where neutrinos stream in all directions, yet with a net
outward flux. Probably one then needs to worry about Eq.~(\ref{eq:EOM1}) as a
full space-time problem, and not as collective evolution along the outward
radial direction, but also not as a simple time evolution of a
quasi-homogeneous system akin to the early universe. Presumably, exponential
growth can happen in both space and time, and moreover may not need to follow
symmetries of the macroscopic system. Moreover, non-forward collisions would
have to be taken into account on some level.

Neutrino flavor evolution under the influence of neutrino-neutrino refraction
remains a challenging subject. As of late, many questions have come into much
clearer focus, yet the full picture has not yet entirely emerged.

\section*{Acknowledgements}

We acknowledge partial support by the Deutsche Forschungsgemeinschaft through
Grant No.\ EXC 153 (Excellence Cluster ``Universe'') and by the European
Union through the Initial Training Network ``Invisibles,'' Grant No.\
PITN-GA-2011-289442.

\section*{References}

%%\bibliographystyle{model1-num-names}
%%\bibliography{sample.bib}


%\begin{raggedright}

%\end{raggedright}

\end{document}